\documentclass[twocolumn,aps,floatfix,showpacs,superscriptaddress]{revtex4}
\usepackage{amsmath,amssymb,amsfonts,eucal,color,graphicx}
\begin{document}
\title{Fragmentation of Random Trees}
\author{Z.~Kalay}
\affiliation{Institute for Integrated Cell-Material Sciences (WPI-iCeMS),
Kyoto University, Yoshida Ushinomiya-cho, Sakyo-ku, 606-8501, Japan}
\author{E.~Ben-Naim}
\affiliation{Theoretical Division and Center for Nonlinear Studies,
Los Alamos National Laboratory, Los Alamos, New Mexico 87545 USA}
\begin{abstract}
  We study fragmentation of a random recursive tree into a forest by
  repeated removal of nodes.  The initial tree consists of $N$ nodes
  and it is generated by sequential addition of nodes with each new
  node attaching to a randomly-selected existing node.  As nodes are
  removed from the tree, one at a time, the tree dissolves into an
  ensemble of separate trees, namely, a forest.  We study statistical
  properties of trees and nodes in this heterogeneous forest, and find
  that the fraction of remaining nodes $m$ characterizes the system in
  the limit $N\to \infty$.  We obtain analytically the size density
  $\phi_s$ of trees of size $s$.  The size density has power-law tail
  $\phi_s\sim s^{-\alpha}$ with exponent
  $\alpha=1+\tfrac{1}{m}$. Therefore, the tail becomes steeper as
  further nodes are removed, and the fragmentation process is unusual
  in that exponent $\alpha$ increases continuously with time. We also
  extend our analysis to the case where nodes are added as well as
  removed, and obtain the asymptotic size density for growing trees.
\end{abstract}
\maketitle

\section{Introduction}

Random trees \cite{hmm,md} underlie a variety of physical processes
including collisions \cite{vv,bkm}, fragmentation \cite{km}, and
fractal aggregation \cite{ym}. These random structures are found in
data storage and retrieval in computer science \cite{dek,jmr,bp,ld}
and they provide a framework for studies in biological evolution
\cite{teh,msw,dekm,mp}.

Previous studies of random trees typically deal with random structures
generated by sequential addition of nodes \cite{hmm,md}. The same
holds for widely-used models of network formation which generally
describe strictly growing networks \cite{dm,ba,dms,krl,mejn}.  Yet, in
many applications including social networks \cite{deb,ptn,cpchbls},
evolutionary trees \cite{msw,dekm}, and technological networks, nodes
may disappear so the network can increase or decrease in size.

\begin{figure}[t]
\includegraphics[width=0.4\textwidth]{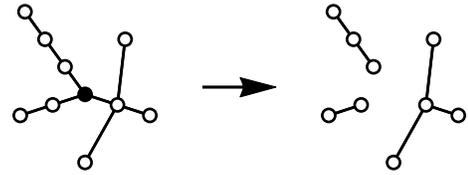}
\caption{The fragmentation process. Removal of the indicated node
  (filled circle) fragments a tree of size $N=10$ (left) into three
  separate trees with sizes $s=4$, $s=3$, and $s=2$ (right). Each
  fragment correspond to a branch of that node.}
\label{fig-frag}
\end{figure}

A number of recent studies of trees formed by addition and removal of
nodes focus on the connectivity of individual nodes and in particular,
the degree distribution \cite{mgn,fc,bk,js,bmrs,sdh}.  Node removal
can cause fragmentation into separate connected components (see Figure
\ref{fig-frag}). Yet, theoretical tools for analyzing connected
components in random structures undergoing fragmentation are limited
\cite{mejn,cpchbls} and statistical properties of groups of connected
nodes in such processes remain largely an open question.

Here, we analyze connected components of a random recursive tree
undergoing fragmentation caused by removal of nodes.  The original
tree is formed by sequential addition of nodes: in each elementary
step one node is added and it is attached to an existing node that is
selected at random. This process is repeated until a tree with $N$
nodes forms. As nodes are removed from the tree, it fragments into
multiple connected components, each having a tree structure.  The
resulting ``forest'' consists of multiple trees (see Figure
\ref{fig-forest}).  Our main goal is to find the size distribution of
connected components in this heterogeneous forest.

As a preliminary step, we establish the fragment-size distribution
when a single node is removed.  This quantity plays the role of a
``kernel'' for the fragmentation process that ensues as nodes are
repeatedly removed from the system.  We use a dynamical formulation
where the number of nodes plays the role of time, and use rate
equations to describe how the fragment-size distribution evolves. We
find the fragment-size distribution analytically as a function of the
fraction of remaining nodes $m$ in the limit $N\to\infty$.  The
density $\phi_s$ of fragments of size $s$ has the algebraic tail
\begin{equation}
\label{phis-tail}
\phi_s \sim s^{-\alpha}\qquad\text{with}\qquad \alpha=1+\frac{1}{m}.
\end{equation}
When an infinitesimal fraction of nodes are removed, the tail is the
broadest, $\phi_s\sim s^{-2}$, but throughout the fragmentation
process, the distribution becomes gradually narrower.  The exponent
$\alpha$ increases monotonically with time and it ultimately diverges
when a finite number of nodes remain.

We also consider growing forests formed by simultaneous addition and
removal of nodes.  In this case, the size distribution is narrower as
it has an exponential tail.

The rest of this article is organized as follows. We first describe
the tree fragmentation process and define the fragment size density
(Section II). In Section III, we consider fragmentation by removal of
a single node and derive the fragment-size density as a function of
size $N$. This quantity allows us to write recursion relations for the
evolution of the size density throughout the fragmentation process. We
obtain a scaling solution where the fraction of remaining nodes $m$
plays the role of a scaling variable (Section IV). In Section V, we
consider the situation where nodes are added and removed at constant
rates, and obtain the leading asymptotic behavior for very large
fragments. We conclude with a summary and a discussion in Sec.~VI. The
Appendix details a few technical derivations.

\section{The fragmentation process} 

We study fragmentation of a random tree through sequential removal of
nodes. The starting point is a random recursive tree
\cite{hmm,md}. This tree is generated by sequential addition of nodes
with each new node attached to a randomly-selected existing node.
Starting with one isolated node, this process repeats until the tree
reaches initial size $N$. The tree has $N-1$ links and hence, it has
no loops.

In each time step, one node is selected at random and it is removed
from the system together with all of the links connected to it.
Hence, the total number of nodes $M$ at time $t$ is simply
\begin{equation}
\label{Mt}
M(t)=N-t.
\end{equation}
At time $t=0$ there are $N$ nodes, and the process ends at time
$t=N-1$ with a single remaining node.

\begin{figure}[t]
\includegraphics[width=0.325\textwidth]{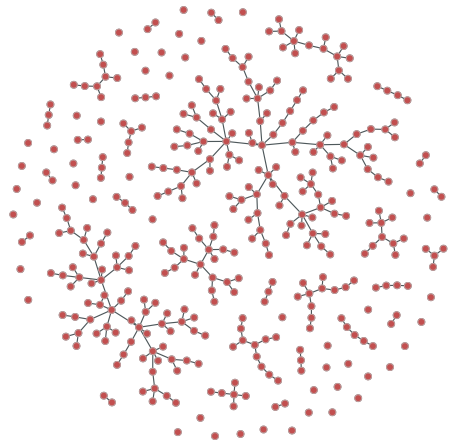}
\caption{The random forest. Shown is a representative example where
  fraction $m=3/4$ of the initial $N=500$ nodes remain.}
\label{fig-forest}
\end{figure}

As nodes are removed from the system, the tree fragments into multiple
connected components. Figure \ref{fig-frag} depicts removal of the
very first node and Figure \ref{fig-forest} shows the resulting forest
after a finite fraction nodes have been removed. Removal of nodes
dissolves the tree into an ensemble of connected components, each
having a tree structure.

The evolving forest is a collection of distinct trees and our primary
goal is to characterize the sizes of trees in this heterogeneous
forest. Let $F_{s,N}(t)$ be the average number of trees of size $s$ at
time $t$ in a system of (initial) size $N$. This quantity corresponds
to an average over infinitely many independent realizations of the
tree generation process and over infinitely many independent
realizations of the node removal process. The initial condition is a
single tree of size $s=N$, $F_{s,N}(0)=\delta_{s,N}$, and the final
state is a single node, $F_{s,N}(N-1)=\delta_{s,1}$.

The total number of trees is deterministic only at the initial and the
final state, but generally, this quantity fluctuates from realization
to realization. In contrast, the total number of nodes \eqref{Mt} is a
deterministic quantity. The number of nodes normalizes the
fragment-size density
\begin{equation}
\label{norm}
\sum_s s\, F_{s,N}(t)=N-t.
\end{equation}
Our main goal is to find the fragment-size density as a function of
time in the limit $N\to\infty$.

\section{The branch-size density}

As a preliminary step, we study $F_{s,N}(1)$, the fragment-size
density after a single node has been removed. We define a branch as
the subtree attached to a node through one of its links.  Figure
\ref{fig-frag} shows that removal of a node with three links and hence
three branches, leads to three fragments.  Therefore, there is a
one-to-one correspondence between the branches of a node and the
fragments generated when that node is removed. Of course, the number
of branches equals the node degree.

Let $P_{s,N}$ be the average number of branches containing $s$ nodes
of a randomly selected node in a random recursive tree of size
$N$. This quantity is equivalent to the fragment-size density at time
$t=1$, that is $P_{s,N}\equiv F_{s,N}(1)$.

The zeroth moment of the branch-size density gives the average number
of branches per node and the first moment gives the total number of
nodes minus one,
\begin{equation}
\label{m01}
\sum_s P_{s,N}=\frac{2(N-1)}{N}, 
\qquad 
\sum_s s\,P_{s,N}=N-1.
\end{equation}
A tree with $N$ nodes has $N-1$ links and every link connects two
branches. Hence, the total number of branches is $2(N-1)$, and the
average number of branches per node is simply $2(N-1)/N$. The second
identity follows from \eqref{norm}. 

The density $P_{s,N}$ satisfies the recursion equation
\begin{eqnarray}
P_{s,N+1} &=&\frac{N}{N+1}\left( \frac{s-1}{N}P_{s-1,N}+
\frac{N-s}{N}P_{s,N} \right)\nonumber\\
&+&\frac{1}{N+1}\left(\delta_{s,1}+\delta_{s,N}\right).
\label{PsN-eq}
\end{eqnarray}
This recursion is subject to the ``initial condition''
\hbox{$P_{s,1}=0$}.  The first two terms account for contributions
from existing branches while the last two terms represent branches
created by the newly added node; hence the respective weights
$\tfrac{1}{N+1}$ and $\tfrac{N}{N+1}$.  When a new node attaches to a
branch of size $s$, the branch size grows by one, that is, $s\to
s+1$. A branch of size $s$ consequently expands with probability
$\tfrac{s}{N}$; otherwise the branch maintains its size with
probability $\tfrac{N-s}{N}$. The last two terms reflect that a new
link generates two new branches, one of size $s=1$ and one of size
$s=N$. By summing \eqref{PsN-eq}, it is possible to check that the
recursion is compatible with the sum rules \eqref{m01}.

Using the recursion relations \eqref{PsN-eq} we can find the
branch-size density for small trees. Starting with $P_{s,1}=0$ we
arrive at
\begin{equation}
\label{PsN-smallN}
P_{s,N}\!=\!
\begin{cases}
\delta_{s,1}& $N=2$, \\
\tfrac{2}{3}\delta_{s,1}+\tfrac{2}{3}\delta_{s,2}&$N=3$,\\
\tfrac{7}{12}\delta_{s,1}+\tfrac{1}{3}\delta_{s,2}+\tfrac{7}{12}\delta_{s,3}&$N=4$, \\
\tfrac{11}{20}\delta_{s,1}+\tfrac{1}{4}\delta_{s,2}+\tfrac{1}{4}\delta_{s,3}+\tfrac{11}{20}\delta_{s,4} &$N=5$.
\end{cases}
\end{equation}
It is also possible to check these expressions by enumerating all
possible tree morphologies for $2\leq N\leq 5$ and removing a
randomly-selected node. Also, the densities listed in 
\eqref{PsN-smallN} satisfy the sum rules \eqref{m01}.

The expressions \eqref{PsN-smallN} suggest that the fragment-size
density is symmetric, $P_{s,N}=P_{N-s,N}$.  This symmetry reflects
that the recursion relation \eqref{PsN-eq} is invariant under the
transformation $s\to N-s$. Moreover, equation \eqref{PsN-smallN}
suggest the general expression
\begin{equation}
\label{PsN}
P_{s,N}=\frac{1}{s(s+1)}+\frac{1}{(N-s)(N+1-s)},
\end{equation}
for $1\leq s \leq N-1$. The small-$s$ expressions adhere to this
general form for all $2\leq N\leq 5$ and all \hbox{$1\leq s\leq
N-1$}. For example, $P_{1,5}=\tfrac{1}{1\cdot 2}+\tfrac{1}{4\cdot 5}$
and $P_{2,5}=\tfrac{1}{2\cdot 3}+\tfrac{1}{3\cdot 4}$. Furthermore,
equation \eqref{PsN} can be justified by induction. When $N=2$, we
have $P_{1,2}=1$ and the expression \eqref{PsN} satisfies the
recursion equation \eqref{PsN-eq} for all $1\leq s\leq N-1$.  We also
note that the first term $\tfrac{1}{s(s+1)}$ in \eqref{PsN} which does
not depend on tree size $N$ coincides with the distribution of
in-component size for a random tree \cite{bk}.

\begin{figure}[t]
\includegraphics[width=0.4\textwidth]{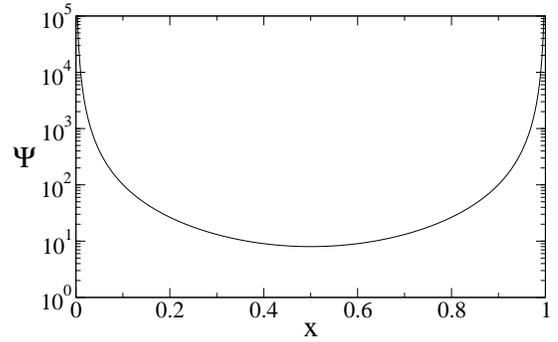}
\caption{The scaling function $\Psi$ versus the scaling variable $x$.}
\label{fig-fx}
\end{figure}

For very large trees, $N\gg 1$, the density \eqref{PsN} adheres to the
scaling form \hbox{$P_{s,N} \simeq (1/N^2)\Psi\left(s/N\right)$}.  In
terms of the normalized branch size $x=s/N$, the scaling function is
(see Figure \ref{fig-fx})
\begin{equation}
\label{Psix}
\Psi(x) = \frac{1}{x^2}+ \frac{1}{(1-x)^2},
\end{equation}
for $0<x<1$ and it has the symmetry $\Psi(x) = \Psi(1-x)$. We note
that the fragment-size density has a power-law tail $F_{s,N}(1)\sim
s^{-2}$ for $s\ll N$ in the limit $N\to \infty$.

\section{The fragment-size density}

We now consider the distribution of fragment size after multiple nodes
have been removed from the system. Importantly, every fragment is
itself a random recursive tree.  Indeed, every fragment is a piece of
the original tree, that is, a subset of connected nodes.  Further,
this subset expands by same growth mechanism that governs the entire
tree. If a new node is attached to this subset, then every one of the
nodes is equally likely to receive this new connection.

This key observation allows us to treat the problem analytically.
Since fragments are equivalent to the initial tree, the outcome of the
very first fragmentation event characterizes all subsequent
fragmentation events. Consequently, the tree fragmentation process
reduces to an ordinary fragmentation process
\cite{cr,zm,es}. Throughout this process, a random tree of size $l$
generates a fragment of size $1\leq s\leq l-1$ with probability
$P_{s,l}$ \cite{rmz,dlb}.  Hence, the tree morphology is entirely
encapsulated by the fragmentation kernel $P_{s,l}$ given in
\eqref{PsN}. 

The fragment-size density evolves according to the linear recursion
equation
\begin{eqnarray}
\label{Fs-eq}
F_s(t+1)=F_s(t)-sf_s(t)+\sum_{l>s} l\,f_l(t)\, P_{s,l}.
\end{eqnarray}
Here, we introduce the rescaled density \hbox{$f_s(t)=F_s(t)/\sum_s
  s\,F_s(t)$}, with the normalization $\sum_s s\,f_s=1$; this
  normalization reflects that the probability a randomly-selected node
  resides in a tree of size $s$ equals $s\,f_s$. Henceforth, the
  dependence on system size is made implicit, $F_s(t)\equiv
  F_{s,N}(t)$.  The recursion equation \eqref{Fs-eq} describes the
  tree fragmentation process.  The loss term equals fragment size
  because the probability the removed node belongs to a tree is
  proportional to tree size. The gain term reflects the fragmentation
  process: trees fragment with probability proportional to their size
  and hence the term $l\,f_l$, while the second quantity $P_{s,l}$ is
  simply the fragmentation kernel. We can verify, using the second
  identity in \eqref{m01}, that the total number of nodes $M=\sum_s
  s\, F_s$ decreases by one in each fragmentation event in agreement 
  with \eqref{Mt}.

Starting with the initial condition $F_s(0)=\delta_{s,N}$, we iterate
the recursion equation \eqref{Fs-eq} once and by construction, recover
the fragmentation kernel $F_s(1)=P_{s,N}$. We now substitute the
expressions \eqref{PsN-smallN} for small $N$ and iterate the recursion
\eqref{Fs-eq} a second time to obtain the fragment-size density once
two nodes are removed,
\begin{equation}
F_s(2)=
\begin{cases}
\delta_{s,1} & $N=3$ \\
\delta_{s,1}+\tfrac{1}{2}\delta_{s,2} & $N=4$ \\
\tfrac{59}{60}\delta_{s,1}+\tfrac{13}{30}\delta_{s,2} +\tfrac{23}{60}\delta_{s,3}& $N=5$. \\
\end{cases}
\end{equation}
These expressions can be manually verified by exact
enumeration. Removal of a second node breaks the symmetry in
\eqref{PsN} because small fragments become more probable at the
expense of large ones.

Our main interest is the behavior in the limit $N\to \infty$.  In this
limit, we can treat time as a continuous variable and convert the
difference equation \eqref{Fs-eq} into the differential equation
\begin{eqnarray}
\label{Fs-eq-cont}
\frac{dF_s}{dt}=-sf_s+\sum_{l>s} l\,f_l\, P_{s,l}.
\end{eqnarray}
The total number of nodes $M=\sum_s F_s$ and the total number of trees
in the forest $T=\sum_s F_s$ obey the differential equations
$dM/dt=-1$ and $dT/dt=1-2T/M$, respectively. These evolution equations
are obtained by summing the rate equations \eqref{Fs-eq} and employing
the first two moments of the fragmentation kernel \eqref{m01}. Using
the initial conditions $M(0)=N$ and $T(0)=1$ we obtain the leading
behavior in the limit $N\to\infty$
\begin{equation}
\label{MT}
M=N\,m,\qquad\text{and}\qquad T=N\,m\,(1-m).
\end{equation}
Here, we introduce the fraction of remaining nodes \hbox{$m=(N-t)/N$}.
As expected, the number of nodes and trees are both proportional to
system size. As a result, the average tree size $\langle s\rangle$ is
given by \hbox{$\langle s\rangle = (1-m)^{-1}$}.

\begin{figure}[t]
\includegraphics[width=0.45\textwidth]{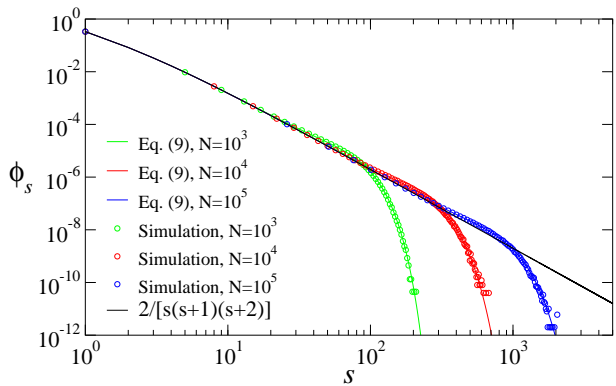}
\caption{The size density $f_s\equiv F_s/M$ versus $s$ for the case
  $m=1/2$.  Shown are: (i) results of numerical iteration of the
  recursion equation \eqref{Fs-eq} for $N=10^3$, $N=10^4$, and
  $N=10^5$ (color lines); (ii) corresponding results of Monte Carlo
  simulations (color circles); and (iii) The limiting distribution
  \eqref{phis} [black line]. Simulation results represent an average
  over $4.5\times 10^4$, $5\times 10^2$, and $10^2$ independent
  realizations with \hbox{$N=10^3$}, $N=10^4$, and $N=10^5$,
  respectively.}
\label{fig-fs}
\end{figure}

The quantities $M$ and $T$ suggest that the fraction of remaining
nodes $m$ characterizes the state of the system in the limit
$N\to\infty$.  Results of numerical integration of the recursion
equation \eqref{Fs-eq} confirm that the fragment-size density depends
only on the fraction of remaining nodes in this limit (see figure
\ref{fig-fs}). Hence, we seek a scaling solution for the fragment-size
density. Formally, the scaling solution $\phi_s$ is defined by
\begin{equation}
\label{scaling}
\phi_s(m)=
\lim_{\substack{N\to\infty\\t\to\infty}}
\frac{F_{s,N}\left(t\right)}{\sum_s s\,F_{s,N}(t)}
\end{equation}
with $m=(N-t)/N$ kept fixed.  The fragment-size density which in
principle depends on two variables, system size $N$ and time $t$,
becomes a function of a single scaling variable, the fraction of
remaining nodes $m$ in the large-size limit. The quantity $s\,\phi_s$
is the probability that a randomly selected node is part of a tree of
size $s$, and accordingly $\sum_s s\,\phi_s=1$.

We now substitute $F_s=M\,\phi_s$ into the rate equation
\eqref{Fs-eq-cont} and introduce the time variable $\tau=\int_0^t dt'
[1/M(t')]$ such that $d\tau/dt=1/M$. This time variable is related to
the fraction of remaining nodes $m=e^{-\tau}$.  With these
transformations, the normalized density $\phi_s\equiv \phi_s(\tau)$
obeys the rate equation
\begin{eqnarray}
\label{phis-eq}
\frac{d\phi_s}{d\tau}\!=\!(1\!-\!s)\phi_s \!+\!\sum_{l>s}\!
\left[\frac{l\,\phi_l}{s(s+1)}\!+\!\frac{l\,\phi_l}{(l-s)(l+1-s)}\right]\!.
\end{eqnarray}
Here, we used the explicit form of $P_{s,l}$.

Results of the numerical integration of \eqref{Fs-eq} suggest that the
tail of the size density is algebraic (see Figure
\ref{fig-fs}). Furthermore, in Appendix A we show that asymptotic
analysis of the rate equation \eqref{phis-eq} yields the power-law
decay \eqref{phis-tail}. In a number of random structures including
networks generated by preferential attachment, power-law tails
correspond to ratios of Gamma functions \cite{krb}. In these contexts,
a ratio of Gamma functions is the discrete analog of an algebraic
function of a continuous variable.  As we show below, such behavior
applies to our fragmentation process.
 
We postulate that the size distribution is a ratio of Gamma functions
\begin{equation}
\label{phis}
\phi_s= (\alpha-2)\frac{\Gamma(s)\Gamma(\alpha)}{\Gamma(s+\alpha)},
\end{equation}
for all $s\geq 1$ with the to-be-determined parameter $\alpha$.  The
prefactor is set by the normalization $\sum_s s\,\phi_s=1$.  In
appendix B, we show that $\phi_s$ in \eqref{phis} satisfies the
``evolution'' equation
\begin{equation}
\label{phis-eq-alpha}
(\alpha\!-\!1)\frac{d\phi_s}{d\alpha}\!=\!
(1\!-\!s)\phi_s\!+\!\sum_{l>s} \!\left[\!\frac{l\,\phi_l}{s(s\!+\!1)}\!+\!\frac{l\,\phi_l}{(l\!-\!s)(l\!+\!1\!-\!s)}\!\right]
\end{equation}
for all $s\geq 1$. This equation describes how $\phi_s(\alpha)$
changes as function of $\alpha$. Since the right-hand sides of
\eqref{phis-eq} and \eqref{phis-eq-alpha} are identical, the original
rate equation \eqref{phis-eq} is satisfied if \hbox{$d\phi_s/d\tau=
(\alpha-1)d\phi_s/d\alpha$}.  Thus, we deduce that \eqref{phis} is a
solution of the evolution equation \eqref{phis-eq} when the parameter
$\alpha$ evolves according to
\begin{equation}
\label{alpha-eq}
\frac{d\alpha}{d\tau}=\alpha-1.
\end{equation}
Together with the initial condition $\alpha(0)=2$ that follows from
\eqref{PsN}, we obtain $\alpha=1+e^\tau$ and find the parameter
$\alpha$ as a function of remaining nodes, 
\begin{equation}
\label{alpha}
\alpha=1+\frac{1}{m}.
\end{equation}
Equations \eqref{phis} and \eqref{alpha} constitute the exact solution
for the scaling function defined in \eqref{scaling}. From this
solution, we can recover the average tree size, $\langle s\rangle =
1/(1-m)$.  The fraction of trees can also be obtained explicitly for
small tree sizes
\begin{equation}
\phi_1=\frac{1-m}{1+m},\qquad \phi_2=\frac{1-m}{1+m}\frac{m}{1+2m}.
\end{equation}

The results \eqref{phis} and \eqref{alpha} establish our main result
announced in \eqref{phis-tail}.  Using the asymptotic behavior
\hbox{$\Gamma(x)/\Gamma(x+a)\to x^{-a}$} as $x\to \infty$ we deduce
$\phi_s\simeq C\,s^{-\alpha}$ with prefactor
$C=(\alpha-2)\Gamma(\alpha)$ and exponent $\alpha$ given in
\eqref{alpha}.  This prefactor vanishes in the limit $\alpha\to 2$
because the number of trees vanishes according to \eqref{MT}.

The exponent $\alpha$ increases monotonically throughout the
fragmentation process. This behavior shows that the tail of the size
density becomes gradually steeper as more nodes are removed.  The size
density \eqref{phis-tail} also has the unusual property that the
exponent governing the tail is time dependent,
$\alpha=1+\tfrac{N}{N-t}$. Power-law tails with {\it fixed} exponents
have been observed in models of fragmentation
\cite{bk1,kbg,kb,pd}. Hence, the tree fragmentation process which has
an unusual nonmonotonic fragmentation kernel \eqref{PsN} also has the
unusual property that the power-law exponent varies continuously with
time.  In many fragmentation processes, the size distribution becomes
universal, namely, it does not evolve with time, once the fragment
size is scaled by the typical fragment size \cite{krb}.  Fragmentation
of random trees does not follow this generic pattern as manifest from
the fact that the exponent \eqref{alpha} is not fixed.
 
The power-law tail \eqref{phis-tail} shows that the forest created by
the node removal process is heterogeneous and includes trees of a
variety of sizes (Figure \ref{fig-forest}). When the size of the
initial tree $N$ is large but finite, the solution \eqref{phis} and
its power-law tail \eqref{phis-tail} hold only for sizes $s\ll s_{\rm
max}$. The size distribution is sharply suppressed beyond this maximal
scale which represents the size of the largest tree in the forest. The
cutoff scale $s_{\rm max}$ can be obtained by the extreme statistics
criterion $1\sim N\sum_{s_{\rm max}}f_s$. From this heuristic
argument, the cutoff grows algebraically with system size
\begin{equation}
\label{smax}
s_{\rm max}\sim N^m.
\end{equation}
Hence, as more nodes are removed, the range of validity of
\eqref{phis} and \eqref{phis-tail} shrinks. For example, when $m=1/2$
then $s_{\rm max}\sim \sqrt{N}$ and as shown in figure \ref{fig-fs},
the rate of convergence toward the $N\to\infty$ limiting behavior
slows down with increasing $N$. In particular, astronomically large
trees are needed to realize the power-law behavior \eqref{phis-tail}
in the limit $m\to 0$.

We verified the theoretical predictions using direct integration of
the recursion equation \eqref{Fs-eq} and using Monte Carlo simulations
of the tree fragmentation process (figure \ref{fig-fs}). The Monte
Carlo simulation results agree with numerical solution of the
recursion equation \eqref{Fs-eq} for finite $N$. This agreement
supports the assertion that fragments remain statistically equivalent
to random recursive trees throughout the fragmentation process. Also,
the numerical results agree with the scaling behavior
\eqref{scaling}. Finally, we confirmed the scaling function
\eqref{phis} for the case $m=1/2$ for which
$\phi_s=2[s(s+1)(s+2)]^{-1}$.

The Monte Carlo simulations were performed by mimicking the tree
creation and fragmentation processes. To generate the initial
configuration, a random recursive tree of size $N$ was
constructed. The tree is formed by sequential addition of nodes. The
attachment probability is uniform such that every existing node is
equally likely to receive a new link. Then, nodes are removed, one at
a time, until $M$ nodes remain. The simulation results presented in
this paper represent an average over multiple independent realizations
of the tree creation and node removal processes.

For completeness, we briefly mention the degree distribution.  We
consider links to be directed and restrict our attention to the
in-degree distribution.  Let $A_k(m)$ be the average number of nodes
with $k$ incoming links once $m$ nodes have been removed. It is well
known that for the random recursive tree, the degree distribution is
exponential, $A_k(0)=N\,2^{-k-1}$ \cite{krb}.  Since the fragments
remain equivalent to a random recursive tree, we expect the degree
distribution to remain exponential. By generalizing similar
calculations in refs.~\cite{mejn,mgn,krb}, it is simple to obtain the
in-degree distribution
\begin{equation}
\label{ak}
A_k=N\,\alpha^{-k-1}.
\end{equation}
The in-degree distribution remains exponential throughout the
fragmentation process and the exponent $\alpha$ governs the
exponential decay of the in-degree distribution.

\section{Addition and Removal of Nodes}

Several recent studies have addressed the situation where nodes can be
added or removed \cite{mgn,fc,bk,js,bmrs,sdh}. In this Section, we
consider the case where nodes are added at constant rate $r$ and
removed (as above) with unit rate. Both processes are completely
random: a newly added node links to a randomly selected node, and
nodes are selected at random for removal. Initially, the system
consists of a single node $M(0)=1$. The number of nodes obeys the rate
equation $dM/dt=r-1$ and hence, it grows steadily with time
\begin{equation}
\label{Mt-add}
M(t)=1+(r-1)t. 
\end{equation}
We restrict our attention to the growing case, $r>1$.  The total
number of trees $T$ is not affected by the addition process, so 
 it evolves according to $dT/dt=1-2T/M$ as above. Solving this
rate equation subject to the initial condition $T(0)=1$, we express
the number of trees as a function of the number of nodes
\begin{equation}
\label{TM-add}
T(M)=\frac{1}{r+1}\,M+\frac{r}{r+1}M^{-2/(r-1)}.
\end{equation}
In the long-time limit, the average tree-size does not depend on time,
$\langle s\rangle \to r+1$ as $t\to\infty$. The second term in
\eqref{TM-add} is negligible in this limit, and we thus conclude that
statistical properties of the forest are characterized by the
parameter $r$ alone.

It is straightforward to generalize the evolution equation
\eqref{Fs-eq-cont} for $F_s(t)$ the average number of trees of size
$s$ at time $t$,
\begin{eqnarray}
\label{Fs-eq-add}
\frac{dF_s}{dt}= r\left[(s\!-\!1)f_{s-1}\!-\!sf_s\right]\!-\!sf_s(t)\!+\!\sum_{l>s} l\,f_l(t)\, P_{s,l}.
\end{eqnarray}
The initial condition is $F_s(0)=\delta_{s,1}$. The first two terms
characterize changes due to node addition and simply reflect that the
probability a tree expands by addition of a new node is proportional
to its size. In writing \eqref{Fs-eq-add} we assume that fragments are
statistically equivalent to a random recursive tree.  By summing this
rate equation, we can verify $dM/dt=r-1$.

Since the average size approaches a constant, we expect that the
normalized size density $F_s/M$ approaches a steady state. Hence, we
define the limiting size-density $\phi_s=\lim_{t\to \infty}
F_s(t)/\sum_s s\,F_s(t)$.  From the evolution equation
\eqref{Fs-eq-add} we deduce that this limiting distribution obeys
\begin{eqnarray}
\label{phis-eq-add}
0&=&r\,[(s-1)\phi_{s-1}-(s+1)\phi_s]  \\
&-&(s-1)\,\phi_s
+\sum_{l>s} \left[\frac{l\,\phi_l}{s(s+1)}+\frac{l\,\phi_l}{(l-s)(l+1-s)}\right].
\nonumber 
\end{eqnarray}

We restrict our attention to very large trees.  The simulation results
suggest that the tail of the size density is exponential, although
there is an algebraic correction,
\begin{equation}
\label{phis-tail-add}
\phi_s\sim s^{-\beta}\, \gamma^s\,,
\end{equation}
for $s\gg 1$.  For such a distribution, the first sum in
\eqref{phis-eq-add} is negligible compared with the second sum, and
the leading behavior of the second sum is (see Appendix C)
\begin{equation}
\label{sum2}
\sum_{l>s} \frac{l\,\phi_l}{(l-s)(l+1-s)}\simeq (c_1\,s+c_2)\,\phi_s,
\end{equation}
in the limit $s\to \infty$.  The two constants are
\hbox{$c_1=1+\big(\gamma^{-1}-1\big)\ln (1-\gamma)$} and
\hbox{$c_2=(\beta-1)\left[1+\gamma^{-1}\ln(1-\gamma)\right]$}. By
substituting \eqref{sum2} into \eqref{phis-eq-add}, we obtain the
recursion relation 
\begin{equation}
\label{phis-eq-rec}
\left(r+1-c_1+\frac{r-1-c_2}{s}\right)\phi_s\simeq r\left(1-\frac{1}{s}\right)\phi_{s-1}
\end{equation}
that applies for very large sizes, $s\to\infty$.  By comparing the two
dominant terms, we confirm the leading exponential behavior
$\phi_s\sim \gamma^s$ with \hbox{$(r+1-c_1)\gamma=r$}. This gives
$\gamma=1-e^{-r}$.  Next, we substitute \eqref{phis-tail-add} into
\eqref{phis-eq-rec} and compare the magnitudes of the leading
corrections $\propto s^{-1}$ to find
$\gamma(r-1-c_2)=r(\beta-1)$. Hence, $\beta=r$, and the tail of the
size density decays according to
\begin{equation}
\label{phis-tail-add-1} 
\phi_s\sim s^{-r}\left(1-e^{-r}\right)^s.
\end{equation}
As the addition rate $r$ grows, the power-law becomes steeper but the
exponential becomes shallower. Consequently, the two terms become
comparable over a growing range with the cross-over scale $s_*\sim
e^{r}$. When $s\ll s_*$, the powerlaw decay is dominant and only when
$s\gg s_*$ does the distribution decay exponentially.  We also note
that in the presence of addition, very large trees become less
probable and the forest becomes less heterogeneous.

\begin{figure}[t]
\includegraphics[width=0.45\textwidth]{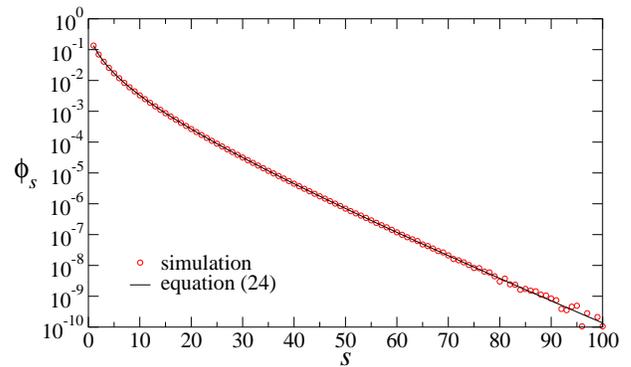}
\caption{The size distribution $\phi_s$ versus $s$ for the case
$r=2$. Shown are results of numerical integration of equation
\eqref{Fs-eq-add}. Also shown are results of Monte Carlo simulations
for $t=10^7$, averaged over $8.5\times 10^3$ independent runs.}
\label{fig-fs-ss}
\end{figure}

We numerically simulated the node addition and removal process. In
each simulation step, with probability $\tfrac{r}{r+1}$ a new node is
added and with probability $\tfrac{1}{r+1}$ one node is deleted. Time
is augmented as follows, $t\to t+\tfrac{1}{1+r}$ after each step.  When a
single node remains, deletion is prohibited.  The Monte Carlo
simulation results confirm that the rate equation \eqref{phis-eq-add}
yields the exact size distribution (Figure \ref{fig-fs-ss}). Hence,
even in the presence of node addition, fragments remain statistically
equivalent to a random recursive tree.

\section{Conclusions}

In summary, we studied fragmentation of a single random tree by
sequential removal of nodes. The emerging random forest is an ensemble
of disconnected random trees of disparate sizes. The size distribution
of trees in this forest has an algebraic tail and the exponent
characterizing this decay increases continuously with the number of
removed nodes.

The original tree expands through sequential addition of nodes and it
dissolves through sequential removal of nodes.  The morphology of the
fragments mirrors that of the parent fragment and thanks to this
feature, knowledge of the very first fragmentation event allows us to
describe the outcome of subsequent fragmentation events. This key
observation enables us to obtain statistical properties of groups of
connected nodes, thereby going beyond characterization of individual
nodes.

We also considered a forest created by addition and removal of nodes
and found that in this case, too, trees remain random throughout the
evolution process. However, the size density becomes narrower and it
has an exponential, rather than algebraic, tail.

It is illuminating to compare fragmentation of random trees with
fragmentation of linear chains of connected nodes where internal nodes
have degree two, except for the end nodes that have degree one. In
this case, the fragmentation kernel is simply $P_{s,l}=2/l$ for $1\leq
s\leq l-1$. Clearly, fragments maintain the linear morphology of the
parent tree.  This case is equivalent to most basic fragmentation
process, the random scission process where a linear rod fragments
repeatedly at a randomly selected location. By substituting the kernel
into \eqref{Fs-eq} and repeating the steps leading to \eqref{phis}, it
is easy to see that the fragment-size density is purely exponential
\cite{krb}
\begin{equation}
\phi_s=(1-m)^2\,m^{s-1},
\end{equation}
where $m$ is again the fraction of remaining nodes.  This example
shows that tree morphology strongly affects the outcome of the
fragmentation process.  The distribution of fragment size can
therefore be used to probe the structure of the fragmented objects.

A number of experiments on fragmentation of solid objects have
reported power-law size distributions with a wide range of exponents
\cite{im,odb,altmo}. It will be indeed interesting to find physical
fragmentation processes \cite{dcaaf,kh,tkch} where the size
distribution becomes steeper with time. Our study deals with discrete
objects where fragment size has a lower bound. A natural extension of
our study is to study fragmentation of more complicated random or
disordered structures.

\acknowledgements 

We acknowledge support by the World Premier International Research
Center (WPI) Initiative of the Ministry of Education, Culture, Sports,
Science and Technology (MEXT) of Japan, and by the US-DOE (grant
DE-AC52-06NA25396).

\appendix
\onecolumngrid\section{Heuristic Derivation of  Equation (1)}

For very large sizes, the first sum in the rate equation \eqref{phis-eq}
is negligible compared with the second.  Let us evaluate the leading
asymptotic behavior of the second sum using the shorthand notation $u(s)=s\,\phi_s$, 
\begin{eqnarray}
\sum_{n=1}^\infty \frac{u(s+n)}{n(n+1)}\simeq  \sum_{n=1}^\infty \frac{u(s)+u'(s)\cdot n}{n(n+1)} 
                                    \simeq  u(s)\sum_{n=1}^\infty \left(\frac{1}{n}-\frac{1}{n+1}\right)+u'(s)\sum_{n=1}^s \frac{1}{n}
                                    \simeq  u(s)+(\ln s)u'(s).
\end{eqnarray}
This heuristic derivation assumes that the size density decays
sufficiently slowly (see also Appendix C).  Hence, the evolution
equation for the size density becomes
\begin{eqnarray}
\frac{\partial \phi_s}{\partial \tau}\simeq (\ln s)\frac{\partial}{\partial s}\left(s\,\phi_s\right).
\end{eqnarray}
This equation gives the power-law tail $\phi_s\sim s^{-\alpha}$ and
the evolution equation \eqref{alpha-eq}.

\section{Derivation of Equation (16)}

We first write the ratio of Gamma functions \eqref{phis} as a product
\begin{equation}
\label{phis-product}
\phi_s=(s-1)!(\alpha-2)\prod_{k=0}^{s-1} \frac{1}{k+\alpha}.
\end{equation}
Next we differentiate $\phi_s$ with respect to the parameter $\alpha$, 
\begin{equation}
\label{phis-log}
\frac{d\phi_s}{d\alpha}=\left(\frac{1}{\alpha-2}-\sum_{k=0}^{s-1}\frac{1}{k+\alpha}\right)\phi_s.
\end{equation}
Equation \eqref{phis-eq-alpha} follows from \eqref{phis-log} and the following identity
\begin{eqnarray}
\label{phis-identity}
(1-s)\phi_s+\sum_{l>s} \frac{l\,\phi_l}{s(s+1)}+\sum_{l>s} \frac{l\,\phi_l}{(l-s)(l-s+1)}=
(\alpha-1)\left(\frac{1}{\alpha-2}-\sum_{k=0}^{s-1}\frac{1}{k+\alpha}\right) \phi_s.
\end{eqnarray}
This identity follows from sums involving the Gamma function. 
The first sum is evaluated as follows 
\begin{eqnarray}
\frac{1}{s(s+1)}\sum_{n=1}^{\infty} \frac{(s+n)!}{\Gamma(s+n+\alpha)} 
&=&\frac{(s+1)!}{s(s+1)\Gamma(s+1+\alpha)}
\left[ 1 + \frac{s+2}{s+1+\alpha} + \frac{(s+2)(s+3)}{(s+1+\alpha)(s+2+\alpha)} + \cdots \right] \nonumber \\
&=& \frac{(s-1)!}{\Gamma(s+1+\alpha)}\, {}_{2} F_{1} (1, s+2; s+1+\alpha; 1) \nonumber\\
&=&\frac{1}{(\alpha-2)}\frac{(s-1)!}{\Gamma(s+\alpha)}.
\end{eqnarray}
In the last step, we used the identity ${}_{2} F_{1} (a, b; c; 1) =
[\Gamma(c) \Gamma(c-a-b)]/[\Gamma(c-a)\Gamma(c-b)]$ obeyed by the
hypergeometric function \cite{gr}. The second sum is evaluated as follows
\begin{eqnarray}
\sum_{l>s}\frac{l!}{(l-s)(l-s+1)\Gamma(l+\alpha)}
&=&
\sum_{l>s}\frac{l!}{(l-s)\Gamma(l+\alpha)}-\sum_{l>s}\frac{l!}{(l-s+1)\Gamma(l+\alpha)}\nonumber\\
&=&
\frac{(s-1)!}{\Gamma(s+\alpha-1)}\left[\left(\frac{s}{s+\alpha-1}-1\right)\left(\sum_{k=-1}^{s-1}\frac{1}{k+\alpha}\right)+\frac{s+1}{s+\alpha-1}\right]\nonumber\\
&=&
\frac{(s-1)!}{\Gamma(s+\alpha)}\left[s-(\alpha-1)\sum_{k=0}^{s-1}\frac{1}{k+\alpha}\right].
\end{eqnarray}

Here, we employed two identities involving the Gamma function. The
first one is shown as follows,
\begin{eqnarray}
\sum_{n=1}^\infty \frac{(s+n)!}{n\Gamma(s+n+\alpha)}&=& 
\frac{(s+1)!}{\Gamma(s+1+\alpha)} \left[ 1 + \frac{s+2}{2(s+1+\alpha)} + \frac{(s+2)(s+3)}{3(s+1+\alpha)(s+2+\alpha)} + \cdots \right] \nonumber \\
&=& \frac{(s+1)!}{\Gamma(s+1+\alpha)}\, {}_{3} F_{2} (1, 1, s+2; 2, s+1+\alpha; 1)\nonumber \\
&=&\frac{s!}{\Gamma(s+\alpha)}\sum_{k=-1}^{s-1}\frac{1}{k+\alpha}.
\end{eqnarray}
A second identity we used follows from this identity by a simple unit
shift in the summation variable
\begin{eqnarray}
\sum_{n=1}^\infty \frac{(s+n)!}{(n+1)\Gamma(s+n+\alpha)}=\frac{(s-1)!}{\Gamma(s+\alpha-1)}
\left(\sum_{k=-1}^{s-1}\frac{1}{k+\alpha}-\frac{s+1}{s+\alpha-1}\right).
\end{eqnarray}

\section{Derivation of Equation (27)}

To derive \eqref{sum2}, we substitute the exponential times power-law
form \eqref{phis-tail-add} into the second sum in \eqref{phis-eq-add}
and evaluate the two leading terms for large $s$
 \begin{eqnarray}
\sum_{n=1}^\infty \frac{(s+n)^{1-\beta}\gamma^{s+n}}{n(n+1)}
&\simeq & s^{1-\beta}\gamma^s \sum_{n=1}^\infty \frac{(1+n\,s^{-1})^{1-\beta}\gamma^{n}}{n(n+1)}\nonumber \\
&\simeq& s^{-\beta}\gamma^s \left(s\,\sum_{n=1}^\infty \frac{\gamma^{n}}{n(n+1)}+(1-\beta) \sum_{n=1}^\infty \frac{\gamma^{n}}{n+1}\right)\nonumber\\
&\simeq &s^{-\beta}\gamma^s \left\{\left[1+(\gamma^{-1}-1)\ln(1-\gamma)\right]\,s+(\beta-1)\left[1+\gamma^{-1}\ln(1-\gamma)\right]\right\}.
\end{eqnarray}

\end{document}